\newcommand{\manuscript}{0}
\ifnum\manuscript=2
\documentclass[11pt, letterpaper]{article}
\newcommand{\shorttitle}[1]{}
\newcommand{\shortauthors}[1]{}
\newcommand{\received}[1]{}
\newcommand{\altaffiltext}[2]{#1 #2}
\newcommand{\altaffilmark}[1]{\ensuremath{^{\mathrm{#1}}}}
\newcommand{\keywords}[1]{\emph{keywords}---#1}
\newcommand{\acknowledgements}[1]{ }
\newcommand {\apgt} {\ {\raise-.5ex\hbox{$\buildrel>\over\sim$}}\ }
\newcommand {\aplt} {\ {\raise-.5ex\hbox{$\buildrel<\over\sim$}}\ }
\renewcommand{\author}[1]{#1}
\renewcommand{\title}[1]{{\centering {\Large \bf #1}\\ }}
\newcommand{\degree}{\ensuremath{^\circ}}
\newcommand{\fdeg}{\ensuremath{^\circ}}

\usepackage[margin=1.2in]{geometry}
\else\ifnum\manuscript=1
\documentclass[12pt,letterpaper,preprint,flushrt]{aastex}
\else
\documentclass{emulateapj}
\fi\fi
\usepackage{graphicx}        % AASTeX preferred
\usepackage{bm}              % boldface math
\usepackage{natbib}
\usepackage[amssymb]{SIunits}

\newcommand{\arone}{148\,GHz}
\newcommand{\artwo}{218\,GHz}
\newcommand{\arthree}{277\,GHz}

\newcommand{\commentx}[1]{}

\newcommand{\etal}{et al.\,}  % Dude: it's not italicized.  See 15th Chicago
\newcommand{\ra}[3]   % right ascension
   {\makebox[1.5em][r]{#1}\makebox[1.5em][r]{#2} \makebox[2em][r]{#3}}

\newcommand{\hms}[3]  % Write HMS in ApJ style.
   {${#1}^{\mathrm{h}}{#2}^{\mathrm{m}}{#3}^{\mathrm{s}}$}

\newcommand{\hmin}[2]  % Write HM in ApJ style.
   {\ensuremath{{#1}^{\mathrm{h}}{#2}^{\mathrm{m}}}}
   
\newcommand{\hours}[1]  % Write H in ApJ style.
   {\ensuremath{{#1}^{\mathrm{h}}}}

\newcommand{\dms}[3]  % Write DMS in ApJ style.
   {\ensuremath{{#1}\degree{#2}\arcminute{#3}\arcsecond}}

\newcommand{\dm}[2]  % Write DM in ApJ style.
   {\ensuremath{{#1}\degree{#2}\arcminute}}

\newcommand{\ukcmb}  % microKelvin_cmb
           {\ensuremath{\micro \kelvin_\mathrm{cmb}}}

\newcommand{\uk}  % microKelvin
           {\ensuremath{\micro \kelvin}}

\newcommand{\fdeg} % fractional degrees
           {\hbox{$.\!\!^{\circ}$}}

\usepackage{color}

\shorttitle{ACT Instrument}
\shortauthors{D.~Swetz~\etal}

\begin{document}

\title{THE ATACAMA COSMOLOGY TELESCOPE: THE RECEIVER AND INSTRUMENTATION}
\author{
D.~S.~Swetz\altaffilmark{1,2},
P.~A.~R.~Ade\altaffilmark{3},
M.~Amiri\altaffilmark{4},
J.~W.~Appel\altaffilmark{5},
E.~S.~Battistelli\altaffilmark{6,4},
B.~Burger\altaffilmark{4},
J.~Chervenak\altaffilmark{7},
M.~J.~Devlin\altaffilmark{1},
S.~R.~Dicker\altaffilmark{1},
W.~B.~Doriese\altaffilmark{2},
R.~D\"{u}nner\altaffilmark{8},
T.~Essinger-Hileman\altaffilmark{5},
R.~P.~Fisher\altaffilmark{5},
J.~W.~Fowler\altaffilmark{5},
M.~Halpern\altaffilmark{4},
M.~Hasselfield\altaffilmark{4},
G.~C.~Hilton\altaffilmark{2},
A.~D.~Hincks\altaffilmark{5},
K.~D.~Irwin\altaffilmark{2},
N.~Jarosik\altaffilmark{5},
M.~Kaul\altaffilmark{1},
J.~Klein\altaffilmark{1},
J.~M.~Lau\altaffilmark{9,10,5},
M.~Limon\altaffilmark{11,1,5},
T.~A.~Marriage\altaffilmark{12},
D.~Marsden\altaffilmark{1},
K.~Martocci\altaffilmark{13,5},
P.~Mauskopf\altaffilmark{3},
H.~Moseley\altaffilmark{7},
C.~B.~Netterfield\altaffilmark{14},
M.~D.~Niemack\altaffilmark{2,5},
M.~R.~Nolta\altaffilmark{15},
L.~A.~Page\altaffilmark{5},
L.~Parker\altaffilmark{5},
S.~T.~Staggs\altaffilmark{5},
O.~Stryzak\altaffilmark{5},
E.~R.~Switzer\altaffilmark{13,5},
R.~Thornton\altaffilmark{1,16},
C.~Tucker\altaffilmark{3},
E.~Wollack\altaffilmark{7},
Y.~Zhao\altaffilmark{5}
}
\altaffiltext{1}{Department of Physics and Astronomy, University of
Pennsylvania, 209 South 33rd Street, Philadelphia, PA, USA 19104}
\altaffiltext{2}{NIST Quantum Devices Group, 325
Broadway Mailcode 817.03, Boulder, CO, USA 80305}
\altaffiltext{3}{School of Physics and Astronomy, Cardiff University, The Parade,
Cardiff, Wales, UK CF24 3AA}
\altaffiltext{4}{Department of Physics and Astronomy, University of
British Columbia, Vancouver, BC, Canada V6T 1Z4}
\altaffiltext{5}{Joseph Henry Laboratories of Physics, Jadwin Hall,
Princeton University, Princeton, NJ, USA 08544}
\altaffiltext{6}{Department of Physics, University of Rome ``La Sapienza'',
Piazzale Aldo Moro 5, I-00185 Rome, Italy}
\altaffiltext{7}{Code 553/665, NASA/Goddard Space Flight Center,
Greenbelt, MD, USA 20771}
\altaffiltext{8}{Departamento de Astronom{\'{i}}a y Astrof{\'{i}}sica,
Facultad de F{\'{i}}sica, Pontific\'{i}a Universidad Cat\'{o}lica,
Casilla 306, Santiago 22, Chile}
\altaffiltext{9}{Kavli Institute for Particle Astrophysics and Cosmology, Stanford
University, Stanford, CA, USA 94305-4085}
\altaffiltext{10}{Department of Physics, Stanford University, Stanford, CA,
USA 94305-4085}
\altaffiltext{11}{Columbia Astrophysics Laboratory, 550 W. 120th St. Mail Code 5247,
New York, NY USA 10027}
\altaffiltext{12}{Department of Astrophysical Sciences, Peyton Hall,
Princeton University, Princeton, NJ USA 08544}
\altaffiltext{13}{Kavli Institute for Cosmological Physics,
5620 South Ellis Ave., Chicago, IL, USA 60637}
\altaffiltext{14}{Department of Physics, University of Toronto,
60 St. George Street, Toronto, ON, Canada M5S 1A7}
\altaffiltext{15}{Canadian Institute for Theoretical Astrophysics, University of
Toronto, Toronto, ON, Canada M5S 3H8}
\altaffiltext{16}{Department of Physics , West Chester University
of Pennsylvania, West Chester, PA, USA 19383}

\begin{abstract}
The Atacama Cosmology Telescope was designed to measure small-scale
anisotropies in the Cosmic Microwave Background and detect galaxy
clusters through the Sunyaev-Zel'dovich effect.  The instrument is
located on Cerro Toco in the Atacama Desert, at an altitude of 5190
meters.  A six-meter off-axis Gregorian telescope feeds a new type of cryogenic
receiver, the Millimeter Bolometer Array Camera.  The receiver
features three 1000-element arrays of transition-edge sensor
bolometers for observations at 148\,GHz, 218\,GHz, and 277\,GHz.  Each detector array is fed by free space mm-wave optics.  Each
frequency band has a field of view of approximately $22\arcmin \times
26\arcmin$.
The telescope was commissioned in 2007 and has completed its third
year of operations.  We discuss the major components of the
telescope, camera, and related systems, and summarize the instrument
performance.
\end{abstract}
\maketitle

\section{Introduction}
\setcounter{footnote}{0}
\label{sec_introduction}

Measurements of the Cosmic Microwave Background (CMB) provide a wealth
of information about the origin and evolution of the Universe.
Current measurements of the angular power spectrum from tenths of a
degree to all sky have provided estimates of cosmological paramters and have begun to quantify the big bang process (e.g., \citet{komatsu/etal:2009}, \citet{brown/etal:2009}, \citet{chiang/etal:prep}, \citet{sievers/etal:prep}).  Detailed
measurements at arcminute scales
are placing tighter constraints on the standard cosmological
model and probe possible deviations.  These higher resolution measurements also reveal secondary effects on CMB
anisotropy, from, for example, the Sunyaev-Zel'dovich (SZ) effect and gravitational
lensing, which are important for understanding structure formation.

We have built the Atacama Cosmology Telescope (ACT), a custom six-meter telescope, to address these
scientific questions.  ACT features a cryogenic receiver, the Millimeter
Bolometer Array Camera (MBAC), that operates at three frequencies: 148\,GHz, 218\,GHz, and 277\,GHz.  Each frequency band is imaged by a $32 \times 32$
array of transition-edge sensor (TES) bolometers.  The telescope was
commissioned at its site in late 2007, at which time the 148\,GHz channel
was installed in MBAC.  The remaining two frequencies were installed
in June 2008.  Since then the telescope and completed camera have
collected approximately twelve months of data over two observing
seasons.

This paper presents the instrument and
complements a set of ACT papers presenting the first generation of
experimental results. The paper is
organized as follows.  In Section \ref{sec_act} we discuss the optical
design and structure characteristics, site location, and telescope
operation.  Section \ref{sec_mbac} details the design of the MBAC
cryostat, cold optics, and detectors.  Telescope control, data
acquisition, merging, and related topics are discussed in Section
\ref{sec_control}.  Finally, we present the image quality of the system
in  Section \ref{sec_beams}.  Companion
papers that detail the beams and scientific
results are \citet{hincks/etal:prep} and \citet{fowler/etal:prep}.

\section{The Atacama Cosmology Telescope}
\label{sec_act}

\subsection{Telescope Construction and Optics}

The diameter of the six-meter primary reflector was set by the requirement to obtain
arcminute-resolution at the ACT frequencies.  The primary and
two-meter secondary are arranged in an off-axis Gregorian
configuration to give an unobstructed image of the sky.  The primary
focal length was fixed at 5.2\,m.  This results in a compact
arrangement between the primary and secondary reflectors, making
it easier to achieve the fast scanning specifications of the telescope
(Section \ref{sec:scan}).  The design is described in
\citet{fowler/etal:2007}.  The telescope was built by Amec Dynamic
Structures Ltd. (now Empire Dynamic Structures).

\begin{figure*}
\begin{tabular}{cc}

\includegraphics[width=3in]{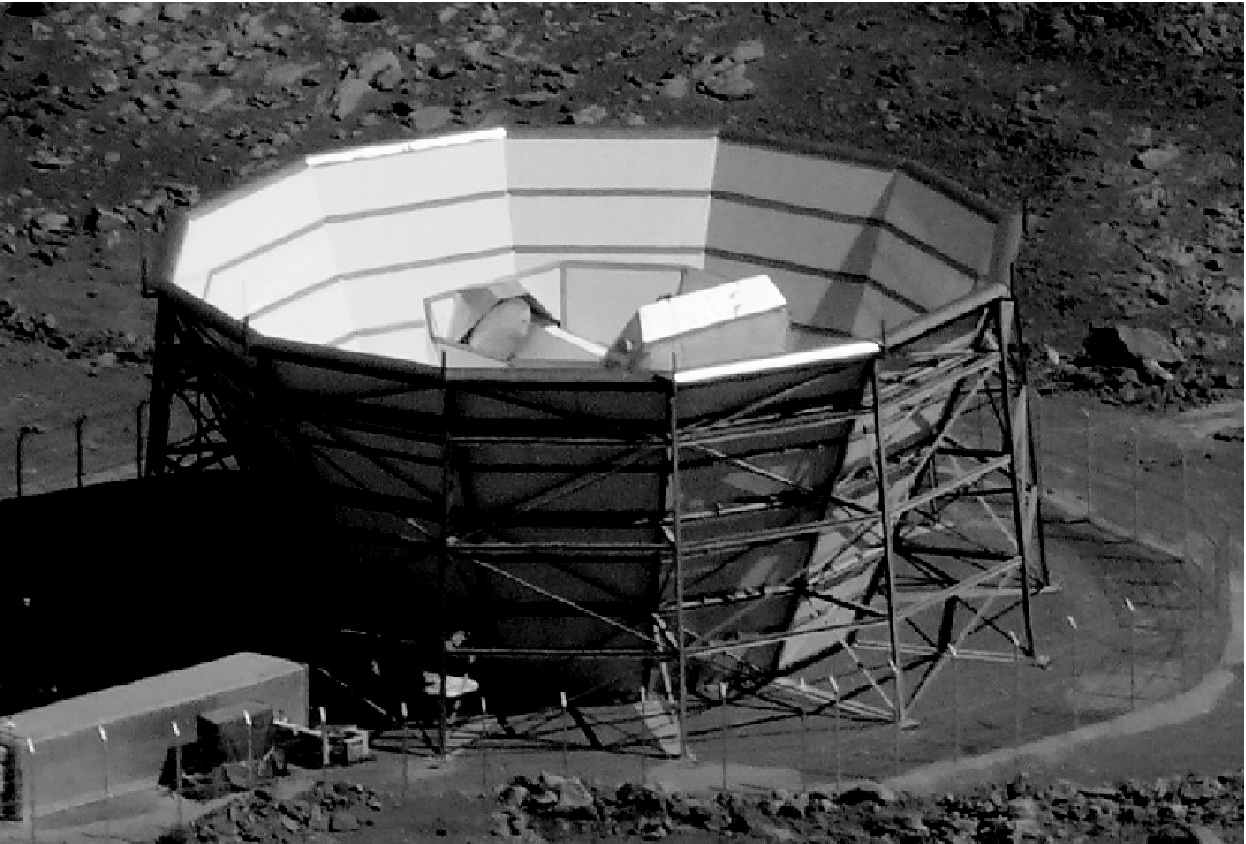}

\hspace{0.5in}
\includegraphics[width=3.5in]{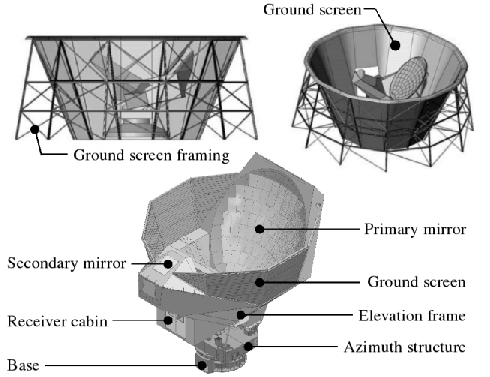}

\end{tabular}

\caption[Mechanical rendering of ACT and its ground
    screens]{\label{fig:telescope}  Picture  {\em (left)} and mechanical rendering  {\em (right)} of ACT and its
    ground screens.  The telescope has a low profile; the full height
    is 12\,m.  The entire upper structure (``Azimuth structure'' and
    above) rotates as a unit.  The surrounding outer ground screen
    shields the telescope from ground emission.  The screen
    also acts as a wind shield.  An inner ground screen mounted on
    the telescope connects the sides of the secondary and primary.
    The primary reflector is 6\,m and is surrounded by a 0.5\,m guard ring
    (Figures courtesy of AMEC Dynamic Structures). }
\end{figure*}

Figure \ref{fig:telescope} shows the major components of the telescope
structure and Table \ref{tab:telescope} the important parameters. To
minimize ground pick-up during scanning, the telescope has two ground
screens.  A large, stationary outer ground screen surrounds the
telescope.  A second, inner ground screen connects the open sides of
the primary reflector to the secondary reflector, and moves with the
telescope during scanning.  A climate-controlled receiver cabin is
situated underneath the primary and secondary reflectors.  The telescope
was designed to work with MBAC (Section
3), and also to be able to accommodate future receivers.

\begin{table}[tbh]
\centering
\caption{Physical properties of the telescope and optics}
\begin{tabular}{rrrr}
\hline \hline

\multicolumn{2}{c}{\bf Telescope Properties} & \multicolumn{2}{c}{\bf Location} \\

Telescope height & 12 m & Altitude & 5190 m \\

Ground screen height & 13 m & Longitude & $67^{\circ}47^{\prime}15^{\prime \prime}$W\\

Total mass & 52 t & Latitude & $22^{\circ}57^{\prime}31^{\prime \prime}$S\\

Moving structure mass & 40 t  & \nodata & \nodata \\

\hline

\multicolumn{4}{c} {\bf Optics} \\
%\multicolumn{2}{c}{\bf Optics} & \multicolumn{2}{c}{\bf Motion} \\

\hline \hline

f-number\tablenotemark{a} & 2.5 & Azimuth range & $\pm 220^{\circ}$ \\

FOV\tablenotemark{a} & 1 deg$^2$ & Max. az speed & $2^{\circ}\!/$s \\

Primary reflector Dia & 6 m & Max. az acc. & $10^{\circ}\!/$s$^2$ \\

No. primary panels & 71 & Elev range & $30\fdeg5$ -- $60^{\circ}$ \\

Secondary reflector Dia & 2 m & Max. elev speed & $0.2^{\circ}\!/$s \\

No. secondary panels & 11 & \nodata & \nodata \\

\hline

\end{tabular}
\tablenotetext{a}{At telescope Gregorian focus}
\label{tab:telescope}
\end{table}

\begin{figure*}
  \begin{center}
    \includegraphics[width=\linewidth]{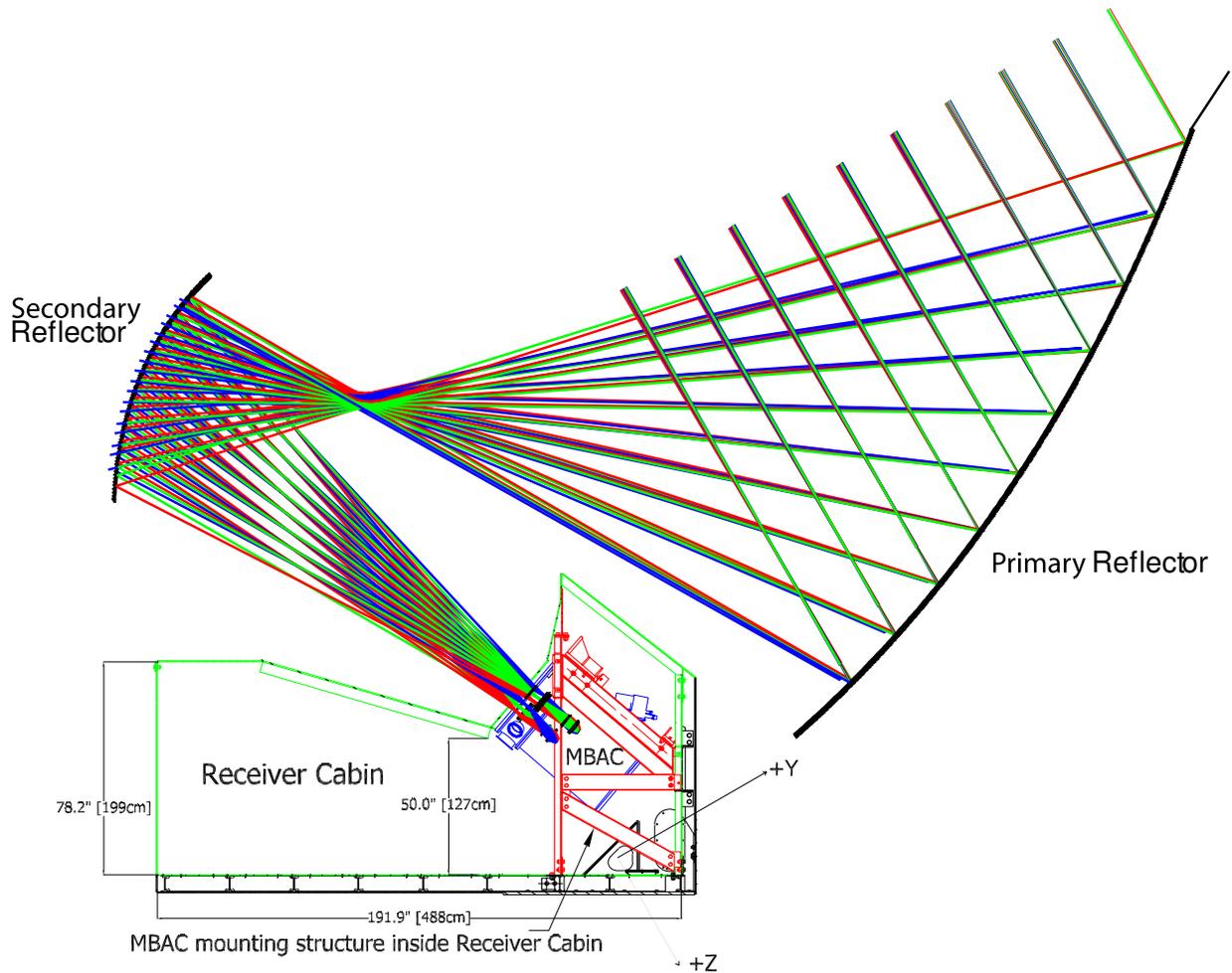}
  \end{center}
  \caption[Ray trace of the off-axis Gregorian optics and MBAC mounted
  in the receiver cabin of the telescope] { \label{fig:cabin} Ray
  trace of ACT's primary and secondary reflectors.  The telescope is an
  off-axis numerically optimized Gregorian.  The rays are traced into
  the MBAC cryostat, mounted on the far right of the receiver
  cabin. The service position (position where the receiver-cabin floor
  is level) is shown, corresponding to a viewing elevation of
  60\degree.  The nominal observing elevation is 50\fdeg5.  The
  rays are traced for the highest (blue), central (green), and lowest
  (red) fields in both the 277\,GHz camera (higher in the cryostat) and the
  218\,GHz camera (lower in the cryostat).  The Figure also shows the
  dimensions and location of the receiver cabin and MBAC mounting
  structure.}
\end{figure*}

Using the Gregorian design as a starting point, the reflector shapes
were numerically optimized to increase the field of view over a classic Gregorian using Code V optical design
software.\footnote{Business address: Optical Research Associates, 3280
East Foothill Blvd., Pasadena, CA, 91107.  Internet URL:
http://www.opticalres.com/.}  At the Gregorian focus before reimaging (Section \ref{sec:optics}), the telescope
achieves a Strehl ratio greater than 0.9 over a 1 square-degree field at 277\,GHz.
Details of the numerical optimization and reflector formulae are given in \citet{fowler/etal:2007}. The telescope
approaches an aplanatic system with no leading-order spherical
aberrations or coma in the focal plane.  Figure \ref{fig:cabin} shows a ray trace through the telescope-camera
system.  The fast focal ratio ($F=2.5$) allows the MBAC window (Section \ref{sec:window}) to be small.

To minimize the beam sizes and maximize the collecting area, 97$\%$ of the primary reflector diameter is illuminated, limited by a cold aperture stop (the ``Lyot stop'').  Spillover at the Lyot stop inside MBAC can
load the detectors with radiation emitted from warm, nearby
structures.  Calculations show that there is a maximum of 0.5\% spillover on the primary reflector and 2\% spillover on the secondary reflector.  To reduce this spillover loading, the primary reflector has a 0.75\,m radial baffle that reduces the primary spillover to $<$\,0.2\%.  However, the spillover on the secondary reflector occurs at larger angles.  While there is a 0.3\,m radial baffle around the secondary, it does little to reduce the effective spillover.  Measurements of the detectors indicate that there is as much as 2--3\% spillover that does not get reflected to the cold sky.  As a result, the secondary baffling is being redesigned to ensure that the majority of the spillover is redirected to the sky in future observations.

\subsection{Site Location, Band Selection, and Logistics}
\label{sec:site}

The ACT site is at an altitude of 5190\,m near the peak of Cerro Toco in the Atacama Desert of northern Chile.  The telescope location provides visibility to approximately 70\% of the sky.
The high elevation and low precipitable water vapor (PWV) at this location provides excellent millimeter and submillimeter atmospheric transparency and has attracted several other millimeter-wave experiments.  The specific ACT site has also been
used by the TOCO \citep{miller/etal:2003} and Millimeter
INTerferometer \citep{fowler/etal:2005} telescopes.

The ACT bands were selected to discriminate between the SZ, CMB, and point sources. The
bands were also chosen to avoid three large emission features, an
oxygen emission line at 119\,GHz and water emission lines at 183\,GHz and
325\,GHz (e.g, \citet{danese89}).  Using the National Radio Astronomy
Observatory/European Southern Observatory monitoring data and an
atmospheric modeling program developed by \citet{pardo/etal:2001}, the
opacity and Raleigh-Jeans (RJ) brightness temperatures are
extrapolated to the ACT bands.
The level of
emission in the continuum where the ACT bands lie is due to O$_2$ and H$_2$O in comparable parts. Thus, the transmission and
absorption in the bands is a function of the PWV in the atmosphere.  Seasonal changes in the weather
provide sustained periods with low amounts of water vapor and
naturally set the ACT observing season to April through December when the PWV is
lowest due to the colder weather.

  The ACT site is approximately 50\,km from the town of San
  Pedro de Atacama (altitude $\sim$ 2750\,m), the location of lodging and the main field
  office.  Travel to the site from North America takes approximately one day.  The roads are clear year-round with brief
  periods of inaccessibility due to snow, providing site access throughout the year.  Communication with the site
  is possible through a data link consisting of a 60\,cm parabolic
  antenna in San Pedro and a 1.2\,m parabolic antenna at the site, and
  an Orthogon PTP600 transreceiver pair operating at 5.8\,GHz.\footnote{Orthogon is a subsidiary of Motorola.  For
  more information see www.motorla.com.}  The
  overall communications rate is $\sim$~40\,Mb/s.  Connection to the
  Internet is made in San Pedro; this connection limits the speed of
  communication from North America and the site to 1--2\,Mb/s.

Site electricity is alternately supplied by one of two XJ150 John Deere (Triton Power Generation) diesel generators, with sufficient on-site fuel for 60 days of autonomous operation at $\sim 240$ liters per day.  The generators are rated at 150\,kW (at sea level), but typical consumption during full operation is approximately 25--30\,kW.

\subsection{Scan Strategy}
\label{sec:scan}
The sky is scanned to separate the CMB signal from drifts in the detectors and atmosphere.  Often, this is done by moving the telescope beam on the
sky on timescales faster than the $1/f$ knee of the low frequency
noise but slower than the time constants of the detectors.  For ACT,
the entire 40 metric-ton upper
structure typically scans at 1\fdeg5\,/s in azimuth while holding the elevation
fixed (typically at 50\fdeg5).  Scans are done at two positions,
east and west of an arc between the south celestial pole (SCP) and the
zenith.  As the sky rotates, an annulus around the SCP is mapped out.

This strategy has several benefits for observations from the ground.
Changing the amount of airmass affects both the gain of the system
due to atmospheric absorbtion and the background loading from gray body
emission.  Keeping the elevation fixed ensures the amount of
atmospheric airmass is relatively constant during a scan, preventing a
scan-synchronous atmospheric signal.  Second, by performing scans both
east and west along the SCP-zenith arc, the mapped annulus will be
observed in two different cross-linked orientations.  Cross-linking has
been shown to be important for the removal of scanning-induced
systematic effects such as striping when making maximum likelihood maps
\citep{wright:1996, tegmark97} of the millimeter sky.  Finally, by moving the entire upper
structure of the telescope, including the primary, secondary, and
receiver, the detectors are constantly looking through the same
optics.
This avoids any
scan-synchronous signals such as beam shape, reflector emission, or ground
pick-up that could potentially arise from changing
the optical path.

\subsection{Pointing}
\label{sec:pointing}

The typical telescope scan has an amplitude
between 5 and 7 degrees with constant velocity between turnarounds, when the telescope changes direction at the end of a scan.
Rapid turnarounds at the end points of the scan minimize
time spent not scanning, which maximizes the sky overlap of the three
separate arrays as discussed in Section \ref{sec:optics} (see Figure
\ref{fig:fov}).  The telescope is designed to produce a maximum scan
speed of 2$\degree$/s with a turnaround time of less than 400\,ms
(10$\degree$/s$^2$) while maintaining a pointing error of 6$^{\prime\prime}$
rms from the commanded position in both azimuth and elevation.  The
rapid turnarounds combined with the 40 metric-ton scanning component
of the telescope made the pointing stability during scanning a
considerable challenge.

The telescope motion systems were designed by KUKA, a robotics systems
company.\footnote{Internet URL: http://www.kuka.com.}  The KUKA robot
comprises the motor drivers, an uninterruptible power supply, an
embedded computer with a solid-state drive, and operator console.  The
KUKA robot monitors the telescope position using a pair of
Heidenhain absolute encoders (27-bit, $0.0097^{\prime\prime}$ accuracy)
on the azimuth and elevation axes.  For
down-stream pointing reconstruction, a second set of identical
encoders is read out synchronously with the $\approx$\,$400$\,Hz
bolometer data stream.  A DeviceNet network relays telescope motion
parameters between the KUKA robot
controller and a housekeeping computer (Section \ref{sec:housekeeping}).\footnote{Open DeviceNet Vendors
  Association (ODVA; odva.org).}  KUKA's robot also produces a stream of telescope health
information (its internal encoder and resolver readouts, motor
currents and temperature) which are broadcast via the user datagram
protocol (UDP) and recorded by the housekeeping computer at $50$\,Hz.
Inclinometers are mounted on the telescope base and the rotating
structure to measure tilt, and accelerometers are mounted at four
positions along the center of the primary reflector and one near the
secondary reflector.  The readout of these devices is synchronized to
within 5\,$\micro$s of the detector data \citep{switzer/etal:2008}.  The
elevation encoder readings are well within the pointing requirements
during the constant-velocity portion of the scan.

During science observations, the scan speed is 1\fdeg5\,/s.  The
acceleration at turnaround is reduced so that it takes 800 ms to
change direction.  With these parameters, the telescope stays within
2$\arcsec$ rms of the desired path the entire time, and the sky is observed
through the entire scan.

\subsection{Reflector Alignment}
\label{sec:panels}
\begin{figure}
  \begin{center}
    \includegraphics[width=\linewidth]{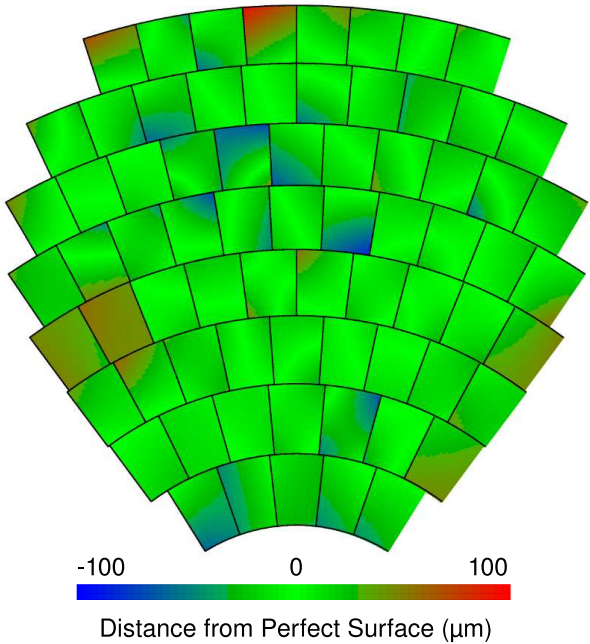}
  \end{center}
  \caption[Primary reflector layout and panel alignment residuals]{
  \label{fig:primary}   Primary reflector layout and final reflector
  alignment.  The primary is composed of 71 approximately rectangular
  panels arranged in eight rows.  The panels are attached to the
  telescope BUS and are aligned using four adjustment screws on the
  panel back-side.  Panel positions are measured using a laser
  tracker.  The residuals of a fit to the reflector's equation give the
  necessary adjustments.  The figure shows the residual after the
  final adjustments for the 2008 observing season were made.  The
  reflector was aligned to better than 30\,$\micro$m rms.}
\end{figure}

The telescope's six-meter primary and two-meter secondary reflectors are
composed of individually machined aluminum panels mounted to a back-up
structure (BUS).  The primary consists of 71 roughly rectangular
panels laid out in eight rows.  The panels comprising each row are
identical, but the curvature of the panels decreases with increasing
height of the BUS (Figure \ref{fig:primary}).  An individual panel
measures $\approx$~$0.65 \times 0.85$\,m and weighs 10\,kg.  The
secondary reflector is assembled by arranging 10 trapezoidal panels,
measuring $\approx 0.35 \times 0.80$\,m, around a $\approx 0.50 \times
0.80$\,m decagonal-shaped central panel.
The panels, manufactured by Forcier Machine Design, were
individually machined to their required surface shape.\footnote{Business
address: 123 Marshall Ave, Petaluma, CA 04052, USA.}
They were measured using a coordinate-measuring machine (CMM) to
have an rms deviation from the expected shape of $\approx$\,3\,$\micro$m.

The backside of each panel of the primary and secondary reflector is
attached to the telescope structure near the four corners of the
panel, and is thus overconstrained.  Threaded mechanisms at these
attachment points allows for manual coarse and fine adjustment of the
position of each panel \citep{woody/etal:2008}.

The loss in forward gain due to panel-to-panel misalignment was estimated
with the Ruze formula \citep{ruze:1966}.  To achieve $\approx$~90\,$\%$ of the optimal forward gain at
our highest frequency of 277\,GHz, a surface rms of 27\,$\micro$m is required assuming a Gaussian distribution of phase errors.  The panel positions are measured
using a laser tracker manufactured by Faro.\footnote{Internet URL:
http://www.faro.com.}  The tracker measures the time of flight of a
laser pulse to determine the distance to a point on the reflector's
surface.  Corner cubes and retroreflectors are used on every panel to
measure alignment.  After three years and six sets of measurements of the primary, the rms has consistently been
in the
25-30\,$\micro$m range, and 10-12\,$\micro$m for the secondary.
Figure \ref{fig:primary}
shows the alignment results obtained prior to the 2009 observing
season.

After the individual panels have been set, the position of the secondary can be
adjusted as a unit using linear actuators mounted on the secondary
frame. This can be performed remotely by the telescope control
software.  The adjusters allow $\pm$1\,cm motion in $y$ and $z$, and
rotations of $\pm$1$\degree$ in azimuth and elevation.  Optimal focus
is achieved by analyzing the detector response to planets at various
positions for the secondary.

\section{The Millimeter Bolometer Array Camera}
\label{sec_mbac}
MBAC is a cryogenic camera designed specifically to meet the
scientific goals of the ACT project.  The design entails three independent
sets of cold
reimaging optics, one for each frequency.  Achieving the sensitivity
goals of the project requires that the detectors operate at 300\,mK.

\subsection{Cold Reimaging Optics}
\label{sec:optics}
\begin{figure*}
  \begin{center}
    \begin{tabular}{c}
      \includegraphics[height=10cm]{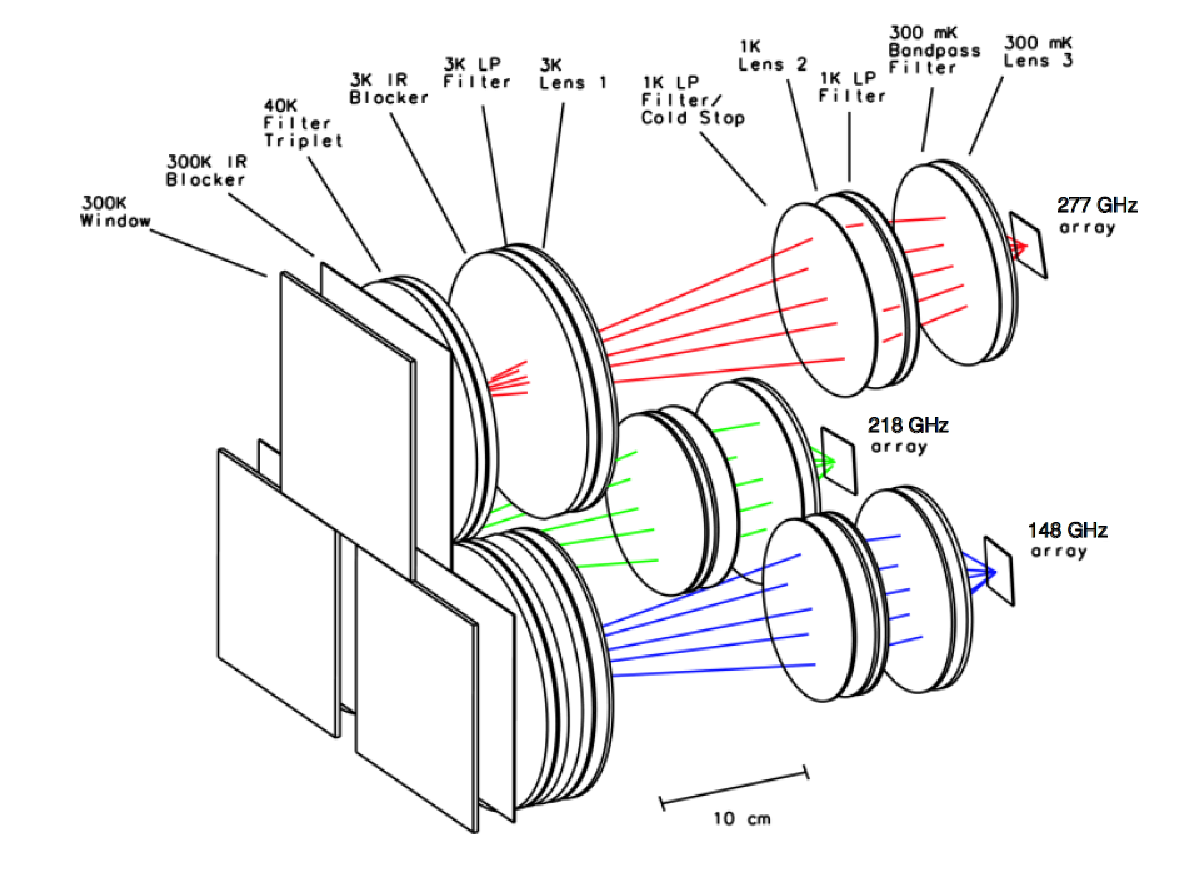}
    \end{tabular}
  \end{center}
  \caption[
    Cold reimaging optics for MBAC.]{\label{fig:optics} Three dimensional model of the
    cold reimaging optics for MBAC.  The optical elements for
    each array are separated into individual optics tubes.  Each
    array has a similar set of optical elements.  The 277\,GHz
    elements and temperatures are labeled.  The lenses are labeled
    Lens 1 to 3, with Lens 1 one being closest to the 300\,K window.
    The low-pass capacitive-mesh filters are labeled LP and the
    band-pass filter as BP.  Infrared blocking filters are labeled
    IR. The temperature  of the components
    decreases moving toward the arrays to reduce the loading, with the
    band-pass filter, the third lens, and arrays all held at 0.3\,K.}
\end{figure*}

The Gregorian telescope produces large field of view suitable for a
several square degree camera.  Three sets of cold optics reimage parts of
this field onto three separate focal plane arrays  (Figure \ref{fig:optics}). Due to the off-axis
Gregorian design, the optimal focal plane is not perpendicular to the
optical axis.  Due to the non-telentric, off-axis design,
the cold optics are held at compound angles inside the cryostat.
The specifications required a diffraction-limited field of view for
each $22\arcmin \times
26\arcmin$ array.  Each set of
reimaging optics has a cold aperture stop that defines the
illumination of the secondary and primary reflectors.

The three optics tube design has several advantages.  Wide-band
anti-reflective (AR) coatings can be difficult to produce, optimizing over a
small band leads to higher transmission through the optical elements.
The optical components at a given frequency are on-axis and
cylindrically symmetric.  This geometry allows the optical elements to be
placed in an optics tube that provides shielding for the arrays
against stray radiation, and leads to a simpler mechanical design.
Each optics tube was built separately, allowing for greater flexibility.
Several aspects of the camera design were based on the MUSTANG instrument on the Green Bank Telescope \citep{dicker/etal:2008} and on the prototype
instrument for ACT, CCAM \citep{lauthesis:2007, Aboobaker:2006}.

A disadvantage of the design is that the three arrays do not image the
same region of sky simultaneously.  The offset between arrays is
minimized by packing them as closely together as
mechanically possible.  The resulting
triangular configuration places the 277\,GHz camera in the plane of
symmetry of the telescope due to its tighter diffraction requirements,
and the 148 and 218\,GHz cameras below and on either side of the plane
of symmetry.
This arrangement coupled with the ACT scan strategy maximizes the
total sky coverage overlap among the different frequencies.  As the
telescope scans back and forth, most of the scanned area is covered by both
of the bottom two arrays; only in the turnaround regions, chosen to be
short in duration, are the two fields
not overlapping.  Sky rotation moves the
lower observed region through the upper camera when observing in the east (or vice-versa when the
telescope is observing in the west).

All the lenses are made out of high-purity silicon, chosen for its high
index of refraction ($n$ = 3.416) \citep{lamb:1996} and
high thermal conductivity at cryogenic temperatures.  High-purity was
used to reduce absorption loss at millimeter wavelengths.  An AR layer
is added to each lens by coating it on both sides with a thin layer
($\approx$ several hundred $\micro$m, the exact number depending on the frequency) of machined Cirlex ($n$ = 1.84) \citep{lau:2006}.  The
first lens, located just after the Gregorian focus, forms an image of
the primary mirror on the cold aperture stop.  Bare arrays, such as those used in MBAC, accept radiation from all directions.  Illumination is controlled using a cold (1\,K) Lyot stop located after the first lens.  Since the cold stop is not a perfect image of the primary, individual detectors geometrically illuminate approximately 5.6\,m of the primary reflector.

The second and third lenses refocus the sky onto the focal plane arrays.  The optics were designed to
maximize the Strehl ratio over the focal plane.
The placement of the final lens
with respect to the array has the tightest tolerance
of all optical elements, $\approx$ 600\,$\micro$m.\footnote{The
tolerance criteria was defined as the amount a lens needed to be
misplaced to produce a 1\% decrease in the Strehl ratio.}  Hence the final lens and array were designed
as a single mechanical unit at 300\,mK.   The final band-pass
filter sits in front of the final lens and is also cooled to
$\approx$~300\,mK (Fig \ref{fig:optics}).

\begin{figure}
  \begin{center}
    \begin{tabular}{c}
      \includegraphics[width=\linewidth]{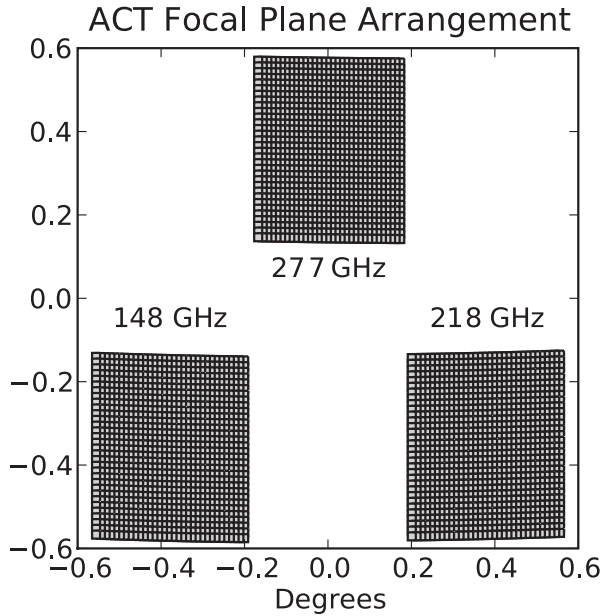}
    \end{tabular}
  \end{center}
  \caption[Detector array locations on the telescope focal plane]{
  \label{fig:fov} The idealized arrangement of the ACT detectors on
  the sky.  This arrangement
on the sky is the reverse of the view when looking into MBAC (Fig. \ref{fig:optics}).}
\end{figure}

The detectors are 1.050~$\times$~1.050\,mm squares, with a
spacing of $1.050 \pm 0.002$\,mm (horizontal) by $1.22 \pm 0.02$\,mm (vertical).  The horizontal error is based on fabrication estimates and the vertical error is based on machine tolerance.  This results in the $32 \times 32$
rectangular detector grid measuring approximately $33.6 \pm 0.1
\times 39.0  \pm 0.1$\,mm, where the errors are estimates from assembly and machining tolerances.  Table \ref{tab:detprop} summarizes their physical and optical
properties.
The numbers for the focal length, primary illumination, and focal ratio vary across the array.  The numbers reported in Table \ref{tab:detprop} are calculated from a model and the ``uncertainties'' are the maximum and minimum range.  The field of view is $ \approx 22\arcmin \times
26\arcmin$ for each array.  In terms of angle on the sky, the detectors
are spaced approximately 1/2$F \lambda$ to 1.1$F \lambda$ going from
the lowest to highest frequencies.
This means that for
148\,GHz, the entire field is fully sampled in a single pointing.
There are no feed horns in the system.  Figure \ref{fig:fov} shows the the relative sky spacing of the
detectors for the three arrays.

\begin{table}[tbh]

\caption{Detector Array Physical and Optical Properties}

\centering
\begin{tabular}{rrr}
\hline\hline
\multicolumn{3}{c}{\bf Detector Physical Properties} \\
\hline
        \nodata & Horizontal & Vertical \\
        Pixel size & 1.050\,mm & 1.050\,mm \\
        Pixel spacing\tablenotemark{a} & 1.050$\pm$0.002\,mm & 1.22$\pm$0.02\,mm \\
        Array configuration & 32 & 32 \\
        Array size\tablenotemark{b} & 33.6$\pm$0.1\,mm & 39.0$\pm$0.1\,mm \\
\hline\hline
\multicolumn{3}{c}{\bf Optical Properties: 148 and 218\,GHz Arrays}\\
\hline
        \nodata & Horizontal & Vertical \\
        Effective focal length\tablenotemark{c} & 5.17$\pm$0.12\,m & 5.03$\pm$0.13\,m \\
        Primary illumination\tablenotemark{c} & 5.60$\pm$0.21\,m & 5.59$\pm$0.23\,m \\
        Focal ratio\tablenotemark{c} & 0.93$\pm$0.02  & 0.90$\pm$0.02 \\
        Detector spacing & 41$\pm$4$''$ & 50$\pm$4$''$ \\
\hline\hline
\multicolumn{3}{c}{\bf Optical Properties: 277\,GHz Array}\\
\hline
        \nodata & Horizontal & Vertical \\
        Effective focal length\tablenotemark{c} & 5.41$\pm$0.11\,m & 5.06$\pm$0.12\,m \\
        Primary illumination\tablenotemark{c} & 5.48$\pm$0.19\,m & 5.50$\pm$0.20\,m \\
        Focal ratio\tablenotemark{c} & 0.98$\pm$0.04  & 0.91$\pm$0.04 \\
        Detector spacing & 40$\pm$4$''$ & 49$\pm$4$''$ \\
\hline
\end{tabular}
\tablenotetext{a}{Horizontal error estimated from fabrication tolerances.  Vertical error estimated from machining tolerances.}
\tablenotetext{b}{Errors are estimated from machining and assembly tolerances.}
\tablenotetext{c}{The number varies across the array and the ``$\pm"$ indicates the maximum and minimum of the range.}
\label{tab:detprop}
\end{table}

Calculations using optics design software predict Strehl ratios greater than 0.97,
0.94, and 0.96 across the entire focal plane region, with average
Strehl ratios of 0.99, 0.98 and 0.98 for the 148, 218, and 277\,GHz
cameras, respectively.

\subsection{Vacuum Windows}
\label{sec:window}

The vacuum window location was chosen near the Gregorian focus of the telescope
to minimize the window size, which in turn reduces the load
on the cryogenic stages.
A separate window was used for each of the
three frequencies, each with an optimized AR coating.  The close
spacing of the optics dictated a rectangular window shape, which takes
advantage of the image shape of the array at the window. The vacuum
windows are made of 4~mm-thick Ultra High Molecular Weight
Polyethylene (UHMWPE) with an index of refraction of $\sim$~1.53 at
160\,GHz \citep{lamb:1996}.
The AR coatings consist of bonded
expanded-Teflon sheets glued to the front and back surfaces of the UHWPE.
Measurements of the MBAC windows
using a Fourier transform spectrometer (FTS) show transmissions of over 96\%
for the 148\,GHz and 218\,GHz bands and over 93\% in the 277\,GHz
band \citep{swetz:2009}.

\subsection{Filters and Bandpass Measurements}
\label{sec:filters}
The frequency responses of the bolometric detector arrays for MBAC are
defined by the transmission through the optical elements in the
optical path to the detectors.  The frequency transmission band is set
primarily by a series of low-pass (LP) capacitive-mesh filters and a
band-defining edge (BP) filter \citep{ade/etal:2006}.
Multiple filters are used because the LP filters
have transmission leaks at harmonics of their cut-off frequency.  The use
of filters with a range of cut-off frequencies allows subsequent
filters to suppress the out of band leaks from previous filters.
Filters are placed at each successive cryogenic stage along the optical
path to limit loading on the next stage.  They are arranged from
highest frequency cut-off to lowest.  There are also a series of
4\,$\micro$m thick infrared-blocking (IR) reflective filters to reduce
the optical loading on the cold stages \citep{tucker/ade:2006}.  These IR
blocking filters prevent the center of the poorly conducting LP
filters from heating up and consequently reduce the radiative loading
from the windows.  Each optics tube contains 10 filter elements,
including the windows.  Table \ref{tab:filters} lists the nominal temperature,
location, and low-pass cut-off frequency for each of the filters.
Figure \ref{fig:optics} shows all the filter and lenses in
free space.

Each filter's frequency response at room temperature was measured using
a FTS.  A composite
transmission spectrum was created by multiplying together the individually
measured filter responses.
After the filters were installed into
MBAC, FTS measurements were made in the lab prior to shipping to Chile with a different spectrometer on the fully cooled system.  It should be noted that at this time there was one additional element
placed in the optical chain, an anti-reflective (AR) layer, directly in
front of the detectors.

Using the instrument passband (normalized by the peak in the composite passband), we follow the method of \citet{page/etal:2003instr} for calculating the
effective central frequency of broadband sources, including
the CMB and SZ effect.   The effect of these sources on the band center is
to shift it slightly.  The results
are given in Table \ref{tab:bp}.

\begin{table}[tbh]

  \caption[Filter location and cut-off frequency in MBAC]{Filter
  location, temperature, and cut-off frequency of the filters in the
  order they are placed in the optical path.
  }
  \label{tab:filters}
\centering
\begin{tabular}{ccccc}
\hline \hline

Filter Type & Temperature & \multicolumn{3}{c}{Frequency~(GHz)}  \\
& & 148 & 218 & 277 \\
\hline

Window & Ambient & \nodata & \nodata & \nodata   \\

IR blocker & Ambient & \nodata & \nodata & \nodata   \\

IR blocker & 40 K  & \nodata & \nodata & \nodata    \\

Low-pass\tablenotemark{a} & 40 K  & 360 & 360  & 540 \\

IR blocker & 40 K  & \nodata & \nodata & \nodata  \\

IR blocker & 4 K  &  \nodata & \nodata & \nodata  \\

Low-pass\tablenotemark{a} & 4 K  &  270 & 330 & 450 \\

Low-pass\tablenotemark{a} & 1 K  &   210 & 300 & 390 \\

Low-pass\tablenotemark{a} & 1 K  &  186 & 255 & 360 \\

Band-pass\tablenotemark{b} & 300 mK & 148  & 220  & 280    \\

\hline

\hline
\end{tabular}
\tablenotetext{a}{Approximate frequency cut-off listed.}
\tablenotetext{b}{Expected band center listed.}
\end{table}

\begin{table}[tbh]

  \caption[Calculated properties of the MBAC filters]{Properties of the MBAC filters.
  }
  \label{tab:bp}
\centering
\begin{tabular}{llll}
\hline \hline

Array & 148\,(GHz) & 218\,(GHz) & 277\,(GHz) \\
\hline

Comp. Bandcenter\tablenotemark{a}& 146.9 & 217.3 & 273.5 \\

Max. Transmission\tablenotemark{a} & 0.74 & 0.72 & 0.69 \\
\hline
Bandcenter\tablenotemark{b}& 149.2$\pm$3.5 & 219.7$\pm$3.5 & 277.4$\pm$3.5 \\

Bandwidth\tablenotemark{b}& 18.4 & 17.0 & 20.9 \\

Noise Bandwidth\tablenotemark{b} & 27.6 & 25.9 & 34.8 \\

Band FWHM\tablenotemark{b} & 27.0  & 22.2 & 30.3 \\

\hline\hline

\multicolumn{3}{l}{Effective Bandcenters\tablenotemark{c}}&  \\

\hline

Synchrotron\tablenotemark{c} & 147.6 & 217.6 & 274.8 \\

Free-free\tablenotemark{c} & 147.9 & 218.0 & 275.4 \\

Rayleigh-Jeans\tablenotemark{c} & 149.0 & 219.1  & 276.7 \\

Dusty source\tablenotemark{c}  & 149.7 & 219.6 & 277.4 \\

CMB\tablenotemark{c}  & 148.4 & 218.3 & 274.7 \\

SZ effect\tablenotemark{c}  & 146.9 & 220.2 & 277.2 \\

\hline
\hline
\multicolumn{3}{l}{Conversion factors}&  \\

\hline

${\rm \delta T_{CMB}/\delta T_{RJ}}$  & 1.71$\pm$0.04 & 3.02$\pm$0.10 & 5.44$\pm$0.20 \\

${\rm \delta W/\delta T_{RJ}}$ (pW/K) & 0.569$\pm$0.055 & 0.483$\pm$0.055 & 0.699$\pm$0.055 \\

$\Gamma$ ($\mu$K/Jy) & 6826$\pm$350 & 5824$\pm$240 & 4373$\pm$550 \\

\hline
\end{tabular}
\tablenotetext{a}{Based on a composite of room temperature measurements.  The source spectrum has been divided out. The
effect of the lenses and coupling to the detectors is not included.}
\tablenotetext{b}{Measurements were made with all filters installed in MBAC along with the lenses and the ACT detectors.  This measurement includes neutral density filters in each band and the coupling of the FTS to MBAC. The neutral density filters are not present when observing the sky.  Three measures of the width of the band are given. The uncertainties are obtained from a combination of the rms of the measurements and an estimate of the systematic error.  The source spectrum has been divided out of the measured response.}
\tablenotetext{c}{Effective band centers for synchrotron emission ($\alpha= -0.7$), free-free emission ($\alpha= -0.1$), Rayleigh-Jeans emission ($\alpha= 2.0$), dusty source emission ($\alpha= 3.5$). These values are based on the average of the
computed composite of all filters and the measured response in MBAC. The uncertainty on all values is 3.5 GHz.  For example,
for the response to the CMB we use 148 GHz, 218 GHz in AR1 and AR2. Since the CMB will be difficult to detect in AR3 we use the RJ band center of 277 GHz.}
\end{table}

\subsection{Detectors}
\label{sec:dets}

MBAC has three 1024-element ($32 \times 32$) detector arrays, one for
each frequency.  The focal plane is filled by a tiling of free-standing bolometric sensors with square absorbers.  The array elements %consist of
are pop-up TES bolometers \citep{benford/etal:2003, Li:1999},
fabricated in the Detector Development Lab at NASA Goddard Space
Flight Center.  Each 20\,m$\Omega$ TES is voltage-biased with a 1\,m$\Omega$ shunt resistor.  Each sensor array is enclosed in a cooled enclosure and its field of view of the incident radiation is limited by the cold stop.

The bolometers are fabricated in columns of 32 elements
on silicon on insulator (SOI) wafers.  These wafers have a $1.45~\micro$m-thick
silicon membrane separated by a thin layer of oxide from the
$450 \micro$m bulk silicon. Each bolometer is a superconducting MoAu
bilayer with $T_c \approx 500$\,mK atop a $1.05\mbox{ mm}\times 1.05\mbox{
mm}$ square of the silicon membrane.  The rest of the square is
implanted with phosphorus ions that serve as the absorber, with a surface impedance of $\approx 100\,\Omega / \Box$.  The
absorber is coupled to the bulk silicon by four 1\,mm-long legs
etched from the silicon membrane with widths between 5-20\,$\micro$m, depending on
the array.  Two of the legs carry superconducting bias wiring for the
TES. These legs define the thermal conductance $G$ to the thermal
bath, and are mechanically compliant enough so that each can be folded
$90^\circ$ so that the square absorbers are orthogonal to the planes
of bulk silicon.  The bulk silicon is then wire bonded to a silicon circuit
board with superconducting (aluminum) wiring which exits the circuit
board on superconducting (tin plated on copper) flexible circuits via
zero insertion force (ZIF) connectors.  These column assemblies are
stacked to form the $32 \times 32$ array of detectors.

The sensitivities of the 148\,GHz and 218\,GHz arrays
were $\approx\,30\,\uk\,$s$^{1/2}$ and $40\,\uk\,$s$^{1/2}$, respectively
in 2008, in CMB temperature
units.  These include atmospheric effects.
Table~\ref{tab:tesparam} lists some of the basic properties of the
TES detectors with their measured standard deviations for each of the three arrays, including their thermal
conductivities, $G$, their critical temperatures, $T_c$, their dark
noise (NEP), and their time constants, $\tau$.  For a more complete
description of the ACT detectors, see
\citet{marriage/chervenak/doriese:2006, niemack/etal:2008, zhao/etal:2008}.

\begin{table}[htdp]

\caption{Detector Parameters}
\begin{center}
\begin{tabular}{  | l  | l  | l   |l  | l |}
\hline
Array & $G$ &$T_c$ & Dark Noise & $\tau$ \\
 & (pW/K) & (mK) & (W / $\sqrt{\mbox{Hz}}$ ) &  (ms) \\
\hline
148 GHz & $80 \pm 20$ & $510 \pm 20$  & $(6\pm 2)\times 10^{-17}$ & $1.9 \pm 0.2$\\
218 GHz & $120 \pm 30$ & $510 \pm 20$ & $(5 \pm 1)\times 10^{-17}$ &  $2.5 \pm 0.4$\\
277 GHz & $300 \pm 60$ & $500 \pm 30$  & $(7 \pm 1)\times 10^{-17}$ &  $1.6 \pm 0.3$\\
\hline
\end{tabular}
\end{center}
\label{tab:tesparam}
\end{table}

\subsection{Detector Readout and Control}
The individual bolometers are read out in a time-multiplexed fashion
 using Superconducting Quantum Interference Device (SQUID) multiplexers \citep{chervenak/etal:1999}.  Prior to exiting the cryostat,
the bolometer signals are amplified by an array of approximately 100 SQUIDS in series held at 4\,K.  The series array
amplifiers and multiplexers are
controlled and read out by the Multi-Channel Electronics (MCE)
\citep{battistelli/etal:2008}.

Each of the three bands has an independent MCE which is mounted
directly onto the MBAC cryostat.  The MCEs,
their power supplies, and MBAC all reside in the telescope receiver
cabin.  The three MCEs are connected to three storage and control
computers in the equipment room through multimode fiber-optic carriers.  The signals
from each MCE are decoded by a PCI card from Astronomical
Research Cameras, Inc. in each of
the three data acquisition computers.\footnote{Internet URL: http://www.astro-cam.com/.}

The base clock rate of the MCE is
50~MHz.  This is divided down to 100 clock cycles per detector row by
32 detector rows plus one row of dark SQUIDs.  Thus, the native
read-rate of the array is $15.15$\,kHz as it leaves the cryostat.
Nyquist inductors at $0.3$\,K
band-limit the response so the array can be multiplexed with minimal
aliased noise while maintaining stability of the loop
\citep{niemack/etal:2008, niemack:2008}.
To downsample the
$15.15$\,kHz multiplexing rate to $399$~Hz, the MCE applies a 4-pole
Butterworth filter with a rolloff $f_{\rm 3dB} = 122$~Hz to the
feedback stream from each detector.  This filter is efficient to
implement digitally and has a flat passband.  The downsampling to $399$~Hz, which can be obtained by
pulling every 38th sample (at $15.15$\,kHz) from the filter stack, is
synchronized by similar clock counting in the synchronization box.  Each time an MCE receives a system trigger, it packages the output of the $32\times 32 + 1 \times
32$ (32 dark SQUIDs) array and sends it over a fiber optic to a PCI
card on its acquisition computer in the equipement room, where it is
buffered and written to disk.  Additional information that fully
specifies the MCE state in each 10-minute acquisition interval is
written to a text ``run file.''

\subsection{Magnetic Shielding}
\label{sec:cryoperm}
The SQUID multiplexers and amplifiers are sensitive to
changing magnetic fields.  They require magnetic
shielding from both Earth's DC field and AC fields induced by the
telescope motion through Earth's DC field and potential fields such as
those generated from the telescope motors.  The series array
amplifiers are located outside of the optics tubes and are
self-contained units enclosed within their own magnetic shielding.
The SQUID multiplexers, however, are mounted on the silicon cards that make
up each column in the detector array holder.

Because of their proximity to the array it would be difficult to
provide individual magnetic shielding.  The effectiveness of the
magnetic shielding is highly dependent on the shield geometry.  Given
the details of geometry of the detector arrays in the optics tube, it
was found that the best solution was to enclose each optics
tube in magnetic shielding.

\begin{figure*}
  \begin{center}
    \includegraphics[angle = 90, width=7in]{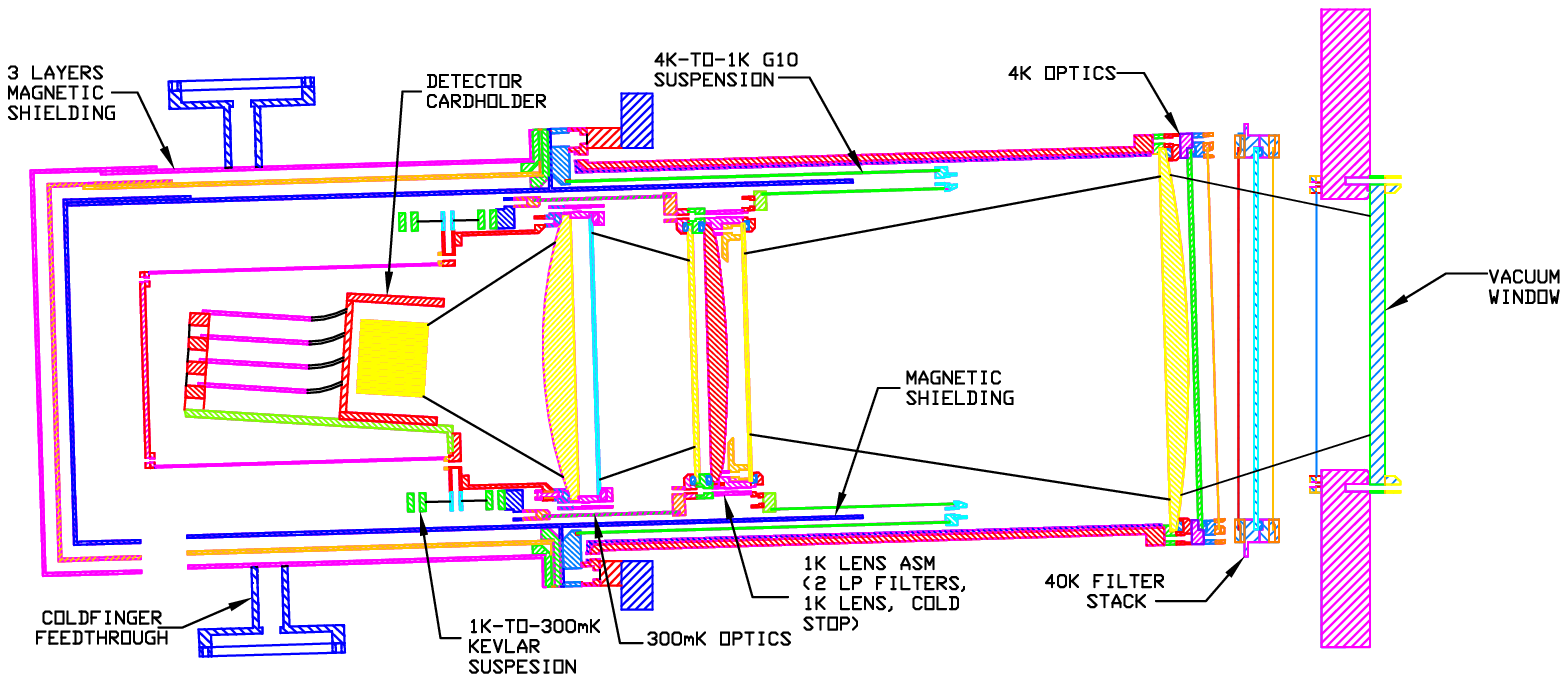}
  \end{center}
  \caption[Cross section that shows some magnetic shielding]{
  \label{fig:magshield}
Cross section of the 148 GHz optics tube.  Elements at five different temperatures, and the structures that separate them,
 are visible.  Multiple layers of
magnetic shielding are used wherever possible, including three layers around
the detector cavity.}
\end{figure*}

Cryoperm-10 is an alloy with a high nickel
  concentration.\footnote{Cryoperm is a trademark of Vacuumschmelze GmbH in
  Hanau, Germany\\Local Distributor: Amuneal Manufactoring
  Corporation, 4737 Darrah St., Philadelphia, PA 19124,
  info@amuneal.com, (800)-755-9843.}  Its composition coupled with a proprietary heat
  treatment give it a high magnetic permeability at cryogenic temperatures.
  To achieve the maximum attenuation, our shielding uses the thickest
  available Cryoperm, 1.5\,mm.  We also use multiple layers
  (Figure~~\ref{fig:magshield}) which,
  given sufficient spacing between them, approaches the limit of
  multiplicative increases in the field attenuation.

Each layer of shielding has different dimensions, ports, and
tubulations, making it difficult to calculate the magnetic field attenuation at
the SQUIDs analytically.  Computer simulations using
Maxwell were employed to apply AC fields to the
geometry of our shields.\footnote{Maxwell is a product of Ansoft.  Internet URL:
  http://www.ansoft.com.}  These simulations estimate the attenuation to be
nearly 40\,dB \citep{thornton/etal:2008}.

\subsection{Cryogenics}
\label{sec:cryogenics}

\begin{figure*}
  \begin{center}
  \includegraphics[width=6in]{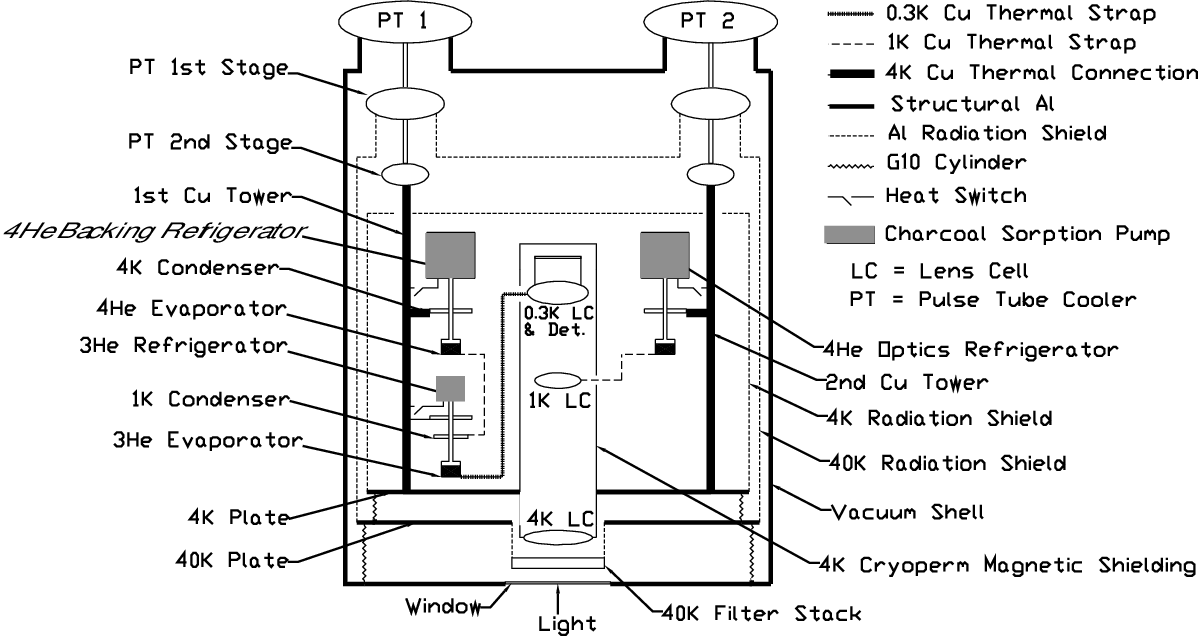}
  \end{center}
  \caption[Schematic of the radiation shielding and thermal connections in MBAC]
  { \label{fig:therm_flow}
Schematic of the radiation shielding and thermal connections in MBAC.  Only a single optics tube is shown, depicting the 148 and
218\,GHz channels.  The 277\,GHz 1\,K optics (not shown) are connected to and cooled by the $^4$He backing refrigerator.
  }
\end{figure*}

The cryogenic design was dictated by a number of requirements.  The
site location makes the use of non-recycled liquid cryogens difficult and expensive; the optics design
requires multiple stages of cooling to reduce detector loading; the
three detector arrays need to be maintained at 300\,mK continuously
for over 12~hours, and the system must recycle in $<$~12~hours; the cryogenics must be stable when the
telescope is scanning.\footnote{The ACT observes from sunset to sunrise; daytime
observations are not possible because solar heating of the telescope
causes deformation of the telescope structure significantly increasing
the $\approx$ 30~$\micro$m\,rms of the primary mirror
\citep{hincks/etal:2008}.}  In order to meet these requirements, MBAC
incorporates three different cooling mechanisms.  A schematic of the
thermal connections in the cryostat is shown in
Figure~\ref{fig:therm_flow}.

The first stage and primary cooling stage is achieved using two pulse
tube cryocoolers.\footnote{Model PT-410 cryorefrigerator from Cryomech.
For more information see www.cryomech.com.}  The first stage of the
pulse tubes cool the 40~K components (Figure \ref{fig:therm_flow}).
The second stage of each pulse tube cooler
(4\,K) is connected to the condenser plate of one of the two custom
$^4$He sorption refrigerators.  The refrigerators have base temperatures of
$\approx$~700\,mK.

The motivation behind two $^4$He refrigators was to use one ``optics''
refrigerator to cool all 1~K optical components, and one ``backing'' refrigerator
to precool the $^3$He refrigerator, the final cooling stage, capable
of reaching 250\,mK.  However, to maximize the hold time of the system,
the cooling of the 280\,GHz 1\,K optics was switched to the $^4$He
``backing'' refrigerator after the first observing season.  At the
coldest stage, the evaporator of the $^3$He refrigerator
cools the 0.3\,K lenses, filters, and detector arrays.  For
more details on the design of the sorption refrigerators,
see \citet{devlin:2004} and \citet{lau:2006}. The $^4$He sorption refrigerators were
measured to have 80\,Joules cooling capacity and the $^3$He refrigerator was measured
to have 5.8\,Joules cooling capacity.

The recycling procedure is completely automated and can be controlled
remotely.  The total recycle time for all three sorption refrigerators
is $\sim$~6~hours.
The detector are thermally regulated
to 313\,mK, slightly above the base temperature of the refrigerator,  to reduce the effects of mechanical heating due to the motion
of the telescope.
This is done with independent servos on each array.

The pulse tube pulses at a frequency of 1.4\,Hz, resulting in an
intrinsic 100\,mK temperature variation at 4\,K and an attendant
mechanical vibration.  Careful selection of thermal connections and
masses prevented thermal oscillations associated with the pulse tube
being detected at either the 1\,K stage or 300\,mK stage.  Mechanical
coupling at the pulse frequency is mitigated by using a compliant
acoustically deadened braided copper rope to attach the pulse-tube
cryocoolers to the cryogenic stages.

The pulse-tube coolers operate at
 maximum capacity only when they are near vertical.  Therefore the
 refrigerators are mounted so that they are vertical when the
 telescope is pointed at 45$\degree$ in elevation, near the middle of
 its range.  Since the telescope also looks at planets for calibration and pointing reconstruction, cryogenic tests
 have been performed with MBAC at both 60$\degree$ and 30$\degree$.
 At these angles, there is little reduction in performance.
A comprehensive description of the
cryogenics is given in \citet{swetz/etal:2008}.

\subsection{Cryostat Mechanical Design}
\label{sec:mechanical}

\begin{figure*}
  \begin{center}
     \includegraphics[width=\linewidth]{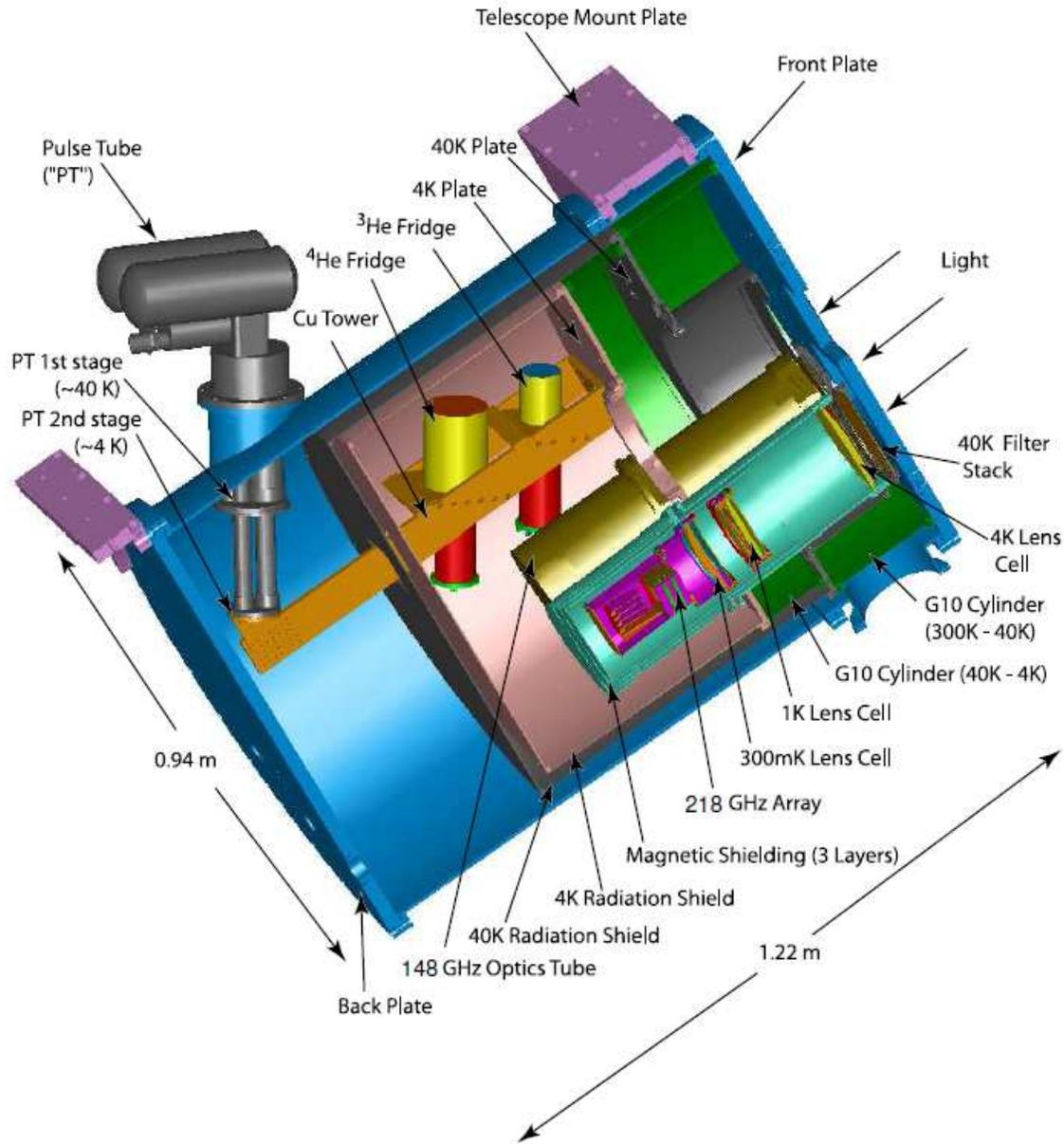}
  \end{center}
  \caption[Cut away of the MBAC cryostat
    showing the location of the internal components.]
  { \label{fig:cutaway} Cut away view of the MBAC cryostat
    showing the location of the internal components.  A cut away view of
    the 218\,GHz optics tube is also shown, giving the location of its
    lenses, filters, and array, which is similar for all three tubes.
    The 277\,GHz optics tube mounts just above the 148 and 218\,GHz
    optics tubes, but is removed for clarity.  The connection of the
    first stage of the pulse tube and its radiation shielding to the
    40\,K plate is also omitted for clarity.}
\end{figure*}

The vacuum shell is a cylinder measuring 0.94~meters in diameter and
1.22~meters long made out of 6.35\,mm (0.25\,inch) thick aluminum.
The shell diameter was dictated by the area of the focal plane, and
the length by the focal length of the reimaging cold optics.  A cut-away illustrating
the major internal components is shown in
Figure~\ref{fig:cutaway}.  The front plate of the shell is 25.4\,mm (1\,inch) thick and serves as the optical bench to which
all of the cold optics are ultimately mounted.  The front plate is
also an integral part of the mounting of MBAC to ACT.  This design
allows all of the cold optics to be rigidly mounted to the telescope
without relying on the rigidity of the cylindrical cryostat vacuum
vessel.

Two additional aluminum plates are attached to the front plate via G-10 cylinders.  The first
plate (``40K plate'') is cooled to $\sim$~40\,K.  The second plate
(``4K plate''), suspended from the 40\,K plate by G-10, is cooled to
$\sim$~4\,K.  A large radiation shield is attached to each plate.  The shields are nested so that the outer 40\,K
shield completely surrounds the inner one 4\,K shield.  The helium refrigerators and optics
tubes are rigidly mounted to the 4K plate and are located between the
4K plate and its corresponding radiation shield.

The vacuum shell is split about 200\,mm back from the front plate.
Removing the back section of the cryostat decouples and removes the
pulse-tube refrigerators from the optics and adsorption refrigerators,
allowing easy access to the cold plates, optics, and detectors.
All of the cabling for the thermometry,
detectors, and the detector readout comes in through ports in the
front vacuum shell section.  These cables are heat sunk at both the
40K and at 4K plates.  An advantage of this design is that it allows
for access to all of the cabling, optics, and detectors without making
or breaking any cable attachments.

\subsection{Optical Support Structures}

To minimize weight, aluminum was generally used to mount optical
elements at 300\,K, 40\,K, and 4\,K.  For the 1\,K and 300\,mK
assemblies, where conductivity is critical and where aluminum is
superconducting (resulting in greatly reduced thermal
conductivity), oxygen-free high-conductivity copper (OFHC)
was generally used.  Another reason for using mostly copper below
$\approx$ 1\,K is potential problems with trapped magnetic flux in
superconducting aluminum alloys.

Each optics tube has a compound wedge at its 4\,K base that holds it at
the proper angle with respect to the telescope beam.  The tubes can be
removed individually, which allowed the 148\,GHz optics tube to be
deployed in MBAC for the 2007 season while the 218 and 277\,GHz optics
tubes were being constructed.

The second lens and cold stop for each optics tube is held at
1\,K.  To reduce the load on our  $^4$He  refrigerator, we
use two concentric and re-entrant G-10 (fiberglass) tubes to connect
the 4\,K optics to the 1\,K optics.

As mentioned in Section \ref{sec:optics}, the third lens/bandpass
and the detector package
were designed to be a single mechanical unit (magenta in Figure \ref{fig:cutaway}).
This OFHC unit is suspended
from the 1\,K optics stack via a Kevlar suspension (not shown in
figure).  The limited cooling capacity at 300\,mK ($\sim$~5.8\,J) and
desired hold time of $>$ 18 hours means the total loading from all
three frequencies must be less than 80\,$\micro$W,
%80\,\microw,
preventing G-10 from being used here.  The
300\,mK assembly is enclosed in a light-tight copper shell which shields
the detector from 4\,K radiation from the surrounding magnetic
shielding.  The total mass of the
300\,mK assembly for each array is approximately 5\,kg.

\subsection{Mechanical Alignment of MBAC to ACT}
MBAC is mounted to ACT at the flange that joins the front plate to the
vacuum cylinder, providing an extremely rigid plane that can be
precisely aligned with the optical axis of the telescope.  This plane
was used as the base for all optical elements inside MBAC, ensuring
that the alignment was independent of variations in the cylindrical
vacuum shell caused by pressure and temperature.  Adjustments permit fine tuning MBAC's position before it is bolted in place.

Figure~\ref{fig:cabin} shows the mating of MBAC in the receiver
cabin.  The width of the receiver cabin is 3.3~meters.  The MBAC enters though doors on the front of the receiver cabin (left side of Figure \ref{fig:cabin}), and passes through the MBAC mounting structure whose width is
$\approx$ 1~meter, where it is then hoisted and bolted into place.  Once the cryostat is bolted into the
telescope receiver cabin, the laser tracker (Section \ref{sec:panels}) is used to locate the front plate in relation to the primary mirror.

\section{Data Acquisition}
\label{sec_control}
\subsection{Overview}

The science data are all written to hard drives on a local
acquisition computer (one for housekeeping and three for the science
cameras).  On each machine, a server broadcasts the data to a central
merger computer which aligns and writes the science data from the
three cameras, telescope and housekeeping systems to a disk at the
telescope site.  Large capacity hard drives are problematic at the ACT site due to the dusty environment and low
atmospheric pressure.  Therefore all of the site drives are contained in individually
pressurized boxes with either
ESATA or USB interfaces.  Because of limited storage space at the site, these merged
data are transmitted to a larger, intermediate, RAID storage node in
San Pedro de Atacama with 6.5\,TB of available storage.  The RAID drives located in San Pedro do not need to be pressurized.  The data can also be served to clients on
real-time data visualization computers either at the
site or the ground station.\footnote{Data are displayed in real time using kst (http://kst.kde.org/).}

To make the large data volume more manageable, we compress the data using a fast, lossless algorithim separately to each channel.\footnote{The compression code {\tt slim} is publicly available at http://slimdata.sourceforge.net/.}  The compression reduces a typical data file to one-third its original size.  The average (compressed) data rate is 80\,GB per night.  The data are copied
from the intermediate RAID storage to external transport disks which
are hand-carried on flights to North America.
Both the housekeeping and camera data
have associated entries in a file database which is periodically
checked to automatically move data from the site computers to the
ground station, onto transport disks, and to confirm that the data
have been properly received (using an md5 checksum) in North America
before deleting the data from site computers.
There is an associated webpage where operators can track the volume and type of
data acquired in real time.

\begin{figure}
\begin{center}
\includegraphics[width=\linewidth]{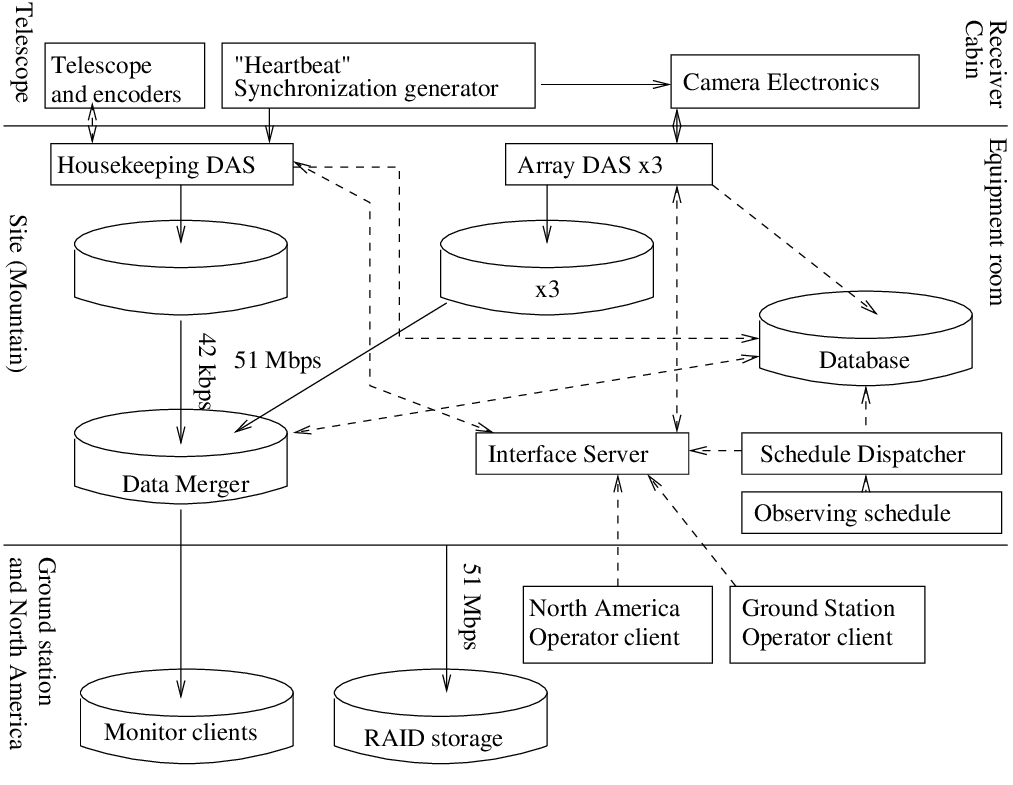}
\end{center}
\caption{ Overview of the ACT data and control systems, split into
telescope (top), on-site (middle), ground station, and North American
systems (bottom).  Solid lines show data streams while dashed lines
show commanding and file information channels.  Cylinders represent
data storage.  We show only one of three detector array acquisition
systems (``Array DAS'') for simplicity.  The ground station RAID array
aggregates data from several machines at the site, so we do not show
it as being connected to a particular node.  To prepare a transport
disk with data to ship back to North America, an operator connects the
drive to the RAID array, and the data are automatically copied over
based on information in the file database. Operators define a schedule
each night and upload it to the schedule dispatcher at the site.  Here
we have only shown a monitor client in the ground station, but the
site also has a system monitor terminal.  }
\label{fig:system_overview}
\end{figure}

Observations with ACT are defined through a schedule file
that specifies an execution time for a task with given parameters.
The schedule is tailored for each observing night and coordinates all
aspects of science operation.  There is no decision structure in the
scheduler itself because of the simplicity of the scan strategy.
Figure~\ref{fig:system_overview} gives a schematic of the acquisition
and commanding system.

\subsection{Timing}

Relative timing to align the camera data, housekeeing and encoders is
determined by a system-wide trigger and associated serial stamp.  An identical
stamp and trigger are passed to each science camera's acquisition
electronics through a fiber optic, and to the housekeeping computer
(in the equipment room) through an RS485 channel in the telescope's
cable wrap.
The system-wide clock rate is defined by
counting down a 25\,MHz clock over 50 cycles per detector row, over 32
rows of detectors plus one row of dark SQUIDs for the array readout,
over 38 array reads per sample trigger.  This gives the final rate of
$\sim 399$~Hz.  The beam-crossing time of a point source in the scan
direction is $\sim 10$~ms, while in the Earth drift direction it is
several seconds.  Thus, the $\sim 399$\,Hz fully samples the beam.
A Meinberg GPS-169 PCI card disciplines the system clock of
the housekeeping computer, with a precision of $<1$~ms to GPS time,
sufficient for astrometry and book-keeping.  Relative timing of
encoders and camera data of $5~\micro {\rm s}$ is achieved.

The synchronization serial stamp from the RS485 channel is
incorporated into the housekeeping data (including the encoders)
through the following chain: 1) in the housekeeping computer, a PCI card
receives a 5\,MHz
biphase signal which encodes serial data stamps at 399\,Hz over RS485
from the synchronization box; 2) these trigger CPU interrupts at 399\,Hz; 3) a timing driver handles these interrupts by polling the
encoders at 399\,Hz; 4) the PCI card clocks down the
serial stamps to 99.7\,Hz, which it uses to poll the housekeeping
acquisition electronics; 5) the housekeeping software then assembles
the housekeeping and encoder time streams, matching serial stamps; 6)
these are written to disk in a flat file format, where for every data
frame there are, for example, four times as many 399\,Hz encoder values as
there are 99.7\,Hz housekeeping data frames.  The camera, encoder, and
housekeeping data are synchronized and stored in the ``dirfile" format, where
one file represents each channel.  For a complete discussion of the ACT timing see \citet{hincks:2009}.

\subsection{Housekeeping readout}
\label{sec:housekeeping}
Housekeeping comprises all electronics and systems other than the
camera and its electronics and the telescope motion control.  The
primary systems within housekeeping are the cryogenic thermal readouts
and controllers, the telescope health readouts, the telescope motion
encoders, and auxiliary monitors.  The telescope's azimuth and
elevation encoder signals are
read by a Heidenhain (model IK220)
PCI card in the housekeeping computer over RS485 from the cable wrap.
Cryogenic housekeeping is read at 99.7\,Hz
and auxiliary channels (such as the mirror temperature) are read from
$1-20$\,Hz asynchronously by a Sensoray 2600 DAQ.  Weather data are
available from an on-site WeatherHawk station and from the APEX
collaboration.\footnote{http://www.apex-telescope.org/weather/index.html.}

\section{Beams}
\label{sec_beams}

Observations of planets were used to determine the positions of
detectors in the focal plane \citep{swetz:2009}, to calibrate
detector response, and to measure beam
profiles for the determination of the window function
\citep{hincks/etal:prep}.  Detector position and beam shape
measurements relied especially on observations of Saturn.

Observations of Saturn were made approximately every second night
during the 2008 season as the planet rose or set through the
ACT observing elevation of 50\fdeg5.  The planet observations were
accomplished using the same fixed-elevation scan strategy and scan
speed as the ACT survey data.

\begin{figure*}[tb]
  \begin{center}
    \begin{tabular}{ccc}
    \includegraphics[scale=.7]{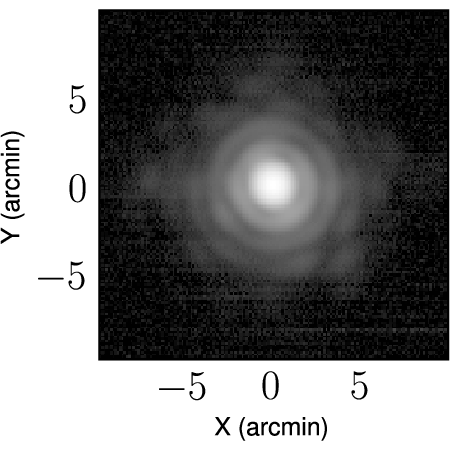}&
    \includegraphics[scale=.7]{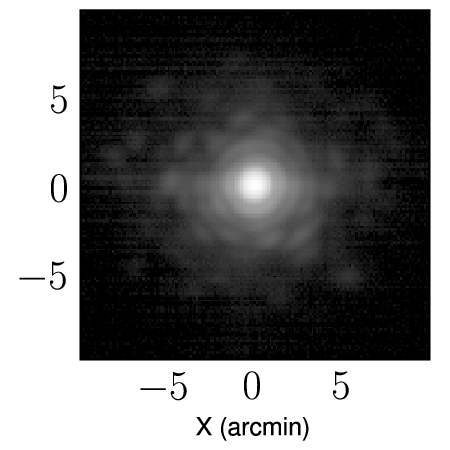}&
    \includegraphics[scale=.7]{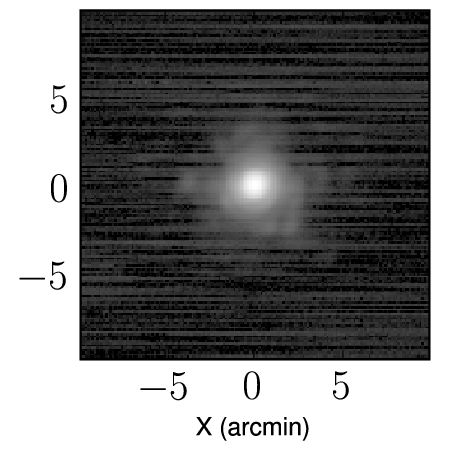}
    \includegraphics[scale=.7]{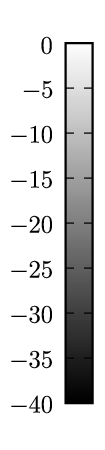}
    \end{tabular}
  \end{center}
  \caption{Beam maps for the \arone\,, \artwo\, and \arthree\,
arrays (from left to right), made from 23, 30, and 13 (respectively)
observations of Saturn.}
\label{fig:beams}
\end{figure*}

The maps are shown in Figure \ref{fig:beams}.
Maximum
likelihood maps were made of each Saturn observation following the method described in \citet{fowler/etal:prep}, but with a less sophisticated approach to the removal of low-frequency atmospheric noise.  For each
map, the detector noise spectra were estimated from the data after
fitting and removing a single common mode.  The solution for each map
was found using a conjugate gradient technique, with a low frequency
common mode removed between each iteration.  Only 5--10 iterations were
necessary to provide adequate convergence.  The maps are made in
tangent plane coordinates, centered at the expected planet peak
position, with the $x$-axis parallel to the scanning direction.

The beam maps in Figure \ref{fig:beams} are the result of co-adding
23, 30, and 13 individual Saturn maps
for 148\,GHz, 218\,GHz, and 277\,GHz, respectively.  (The lower yield of
high quality maps in 277\,GHz is due to relative gain drifts in some
detectors.)  Small, variable offsets in the planet position relative
to the center of each map were removed prior to combining the maps.

Point-spread function quality was assessed by measuring the solid
angle and FWHM of the beam maps in each array.  The
results are summarized in Table \ref{table_beam} and the solid angles are consistent with those presented in \citet{hincks/etal:prep} which used a different map-making method and data subset.  An elliptical Gaussian is fitted
to the peak of the beam profile.  The axis angle is the angle of the
semi-major axis relative to the $x$-axis, increasing
counter-clockwise.\\

\begin{table}[hbt]
  \caption{Summary of Beam Parameters}
    \label{table_beam}
    \begin{center}
\begin{tabular}{lccc}
\hline \hline

& \arone & \artwo & \arthree   \\
\hline
Solid Angle\tablenotemark{a} (nsr) & $217.7 \pm 3.6$ & $117.6 \pm 2.8$ & $97 \pm 12 $ \\
Major FWHM$^{a}$ ($\arcminute$) & $1.401 \pm 0.003$ &
                            $1.012 \pm 0.001$ & $0.891 \pm 0.04$ \\
Minor FWHM$^{a}$ ($\arcminute$) & $1.336 \pm 0.001$ &
                            $0.991 \pm 0.001$ & $0.858 \pm 0.005$ \\
Axis Angle$^{a}$ ($\degree$) & $66 \pm 1$ & $45 \pm 2$ &
                                       $66 \pm 10$ \\
\hline
\end{tabular}
\tablenotetext{a}{The error bars are effective 1-$\sigma$ errors.  The relatively large size of the 277\,GHz error bar is due to the systematic errors from atmospheric modeling and is expected to improve.}
\end{center}
\end{table}

The beam shapes are consistent with an Airy pattern integrated over
the frequency band-width of the filters and non-zero extent of the
detectors.  First side lobe peaks are apparent at approximately the
$-$15 dB level, at radii of roughly 1.7$\arcminute$, 1.2$\arcminute$ and 1.0$\arcminute$.
A simple model of the optics predicts beam FWHM of 1.38$\arcminute$, 1.03$\arcminute$ and 0.88$\arcminute$ for the three arrays \citep{niemack:2008}.  This model was calculated for a single detector near the middle of the array and assumes uniform primary
illumination, a circular aperture, and takes into account the
non-point-like extent of the detectors.  It does not include diffraction analysis or integration of beam variation across the array, while the planet measurements are an average across all detectors in each array.  Despite the simplifications, the measured FWHM are consistent with the model.

\acknowledgements

The ACT project was proposed in 2000 and funded Jan 1, 2004. Many have
contributed to the project since its inception. We especially wish to
thank Asad Aboobaker, Christine Allen, Dominic Benford, Paul Bode,
Kristen Burgess, Angelica de Oliveira-Costa, Peter Hargrave, Amber
Miller, Carl Reintsema, Uros Seljak, Martin Spergel, Johannes Staghun,
Carl Stahle, Max Tegmark, Masao Uehara, and Ed Wishnow. It is a
pleasure to acknowledge Bob Margolis, ACT's project manager. Reed
Plimpton and David Jacobson worked at the telescope during the 2008
season, and Mike McLaren during the 2009 season.  ACT is on the
Chajnantor Science preserve which was made possible by the Chilean
Comisi\'on Nacional de Investigaci\'on Cient\'ifica y Tecnol\'ogica.
We are grateful for the assistance we received at various times from
the ALMA, APEX, ASTE, CBI/QUIET, and Nanten groups. The PWV data come
from the public APEX weather site.  Field operations were based at the
Don Esteban facility run by Astro-Norte. We thank Tom Herbig who
chaired our external advisory board (with Amy Newbury, Charles Alcock,
Walter Gear, Cliff Jackson, and Paul Steinhardt), which helped guide
the project to fruition.  This work was supported by the U.S. National
Science Foundation through awards AST-0408698 for the ACT project, and
PHY-0355328, AST-0707731 and PIRE-0507768. Funding was also provided
by Princeton University and the University of Pennsylvania. AH
received additional support from a Natural Science and Engineering
Research Council of Canada (NSERC) PGS-D scholarship.  PhD thesis based
on ACT can be found at www.physics.princeton.edu/act/papers.html.

We thank AMEC/Dynamic Structures/Empire for their work on the
telescope construction, and Kuka for the motion control system.  We
are very grateful for Bill Dix, Glen Atkinson, and the rest of the
Princeton Physics Department Machine Shop, and Harold Borders at the
University of Pennsylvania Physics Department Machine Shop.  Much of
our data acquisition system was based on the BLAST experiment.  This
paper includes
contributions from a U.S. government agency, and is not subject to copyright.

\pagebreak  % this is needed to keep the figure from the middle of the bib
%\bibliographystyle{plainnat}
%\bibliography{ms}
%\bibliographystyle{astron}

\end{document}